\documentclass[aps,pra,amsfonts,twocolumn,superscriptaddress]{revtex4-1}
\begin{document}
\title{Communication Complexity Reduction from Globally Uncorrelated States}
\author{Marcin Wie\'sniak}\affiliation{Institute for Theoretical Physics and Astrophysics, University of Gda\'nsk, 80-952 Gda\'nsk, Poland}
\begin{abstract}
Bell inequality violating entangled states are the working horse for many potential quantum information processing applications, including secret sharing, cryptographic key distribution and communication complexity reduction in distributed computing. Here we explicitly demonstrate the power of certain multi-qubit states to improve the efficiency of partners in joint computation of some multi-qubit function, despite the fact that there could be no correlations between all distributed particles. It is important to stress that the class of functions that can be computed more efficiently is widened, as compared with the standard Bell inequalities.
\end{abstract}
\maketitle
Entanglement is not only a curious feature of Quantum Mechanics. It is also a resource, which can be useful in specific information processing task. For example, entanglement is not responsible for quantum cryptographic key generation \cite{BB84}, however it allows to verify the presence of an eavesdropper \cite{EKERT}. Also, there are proposals of universal quantum computing, with a promise of speedup with respect to analogous classical protocols, based on multipartite entangled states. Lastly, it has been argued that any instance of violation of a Bell inequality \cite{Bell} leads to the reduction of communication complexity in a certain distributed problem \cite{CCGHZ}, or at least to an increase of the success rate of yielding results \cite{COMCOP,CCPAWL}.

It shall be noted that communication complexity reduction protocoles (CCR) are valid information processing routines. The users have classical bits as an imput and the output is also classical. Hence, they are ready to be applied in the state-of-art computing.

Recently we observe an intensive development in deriving new variants of Bell Inequalities. We are no longer constrained to Werner-Wolf-Weinfurter-\.Zukowski-Brukner (WWW\.ZB) \cite{WW,WZ,ZB} inequalities, which utilize only two settings per observer. Inequalities with more settings were introduced, for example, in Refs. \cite{SLIWA,CHEN,LPZB,WBZ,VERTESI1}. Also, methods to derive Bell Inequalities with correlations between subsets of systems (``subcorrelations'') were presented, for example, in Refs \cite{WNZ,VERTESI2}.

Even though the possibility of increasing the accuracy of distributed computing under limited communication with quantum subcorrelations was observed by Paw\l owski in Ref. \cite{CCPAWL}, an explicit description of the method to attain such a gain is, up to date, missing in literature. This is the aim of this contribution. We want to demonstrate that even globally uncorrelated states might be useful in computational task. The approach described here straightforwardly extends to multisetting inequalities. We will show the class of functions that  can be estimated with the quantum gain and discuss signaling strategies for the partners.

Let us start with recalling the protocol. Partners $A, B, C,...$ are in separate labs (in reality, these could be separate parts of a computer) and at certain point they all receive two pieces of data. One is a bit, $y_i=\pm 1 (i=A,B,C,...)$. The other variables, $x_i$'s, are multi-valued. These values are distributed by a referee (who thus has the complete knowledge of these data and is able to verify the decider's answer), however, we assume that these data could not be known before the protocol has started (e. g., they are randomly generated just before the transmission). There can be correlations between $x_i$'s distributed in individual runs of the task, but each $y_i$ is completely random. Partners can perform any local action, but they are asked to transmit {\em a single bit} to the decider. His task is to calculate a certain sign function $f'(\vec{x},\vec{y})= y_Ay_B...f(\vec{x})=\pm 1$, where $\vec{x}=(x_A,x_B,...)$ and $\vec{y}=(y_A,y_B,...)$. To neutralize the product $y_i$'s in $f'(\vec{x},\vec{y})$, the partners make their messages proportional to $y_i$. If even one of them fails to do so, the final answer is completely random. 

In the classical protocol, the information about the value of $x_i$ may be encoded into the message $m_i$ sent to the decider. In case of bi-valued $x_i$'s, $x_i=1,2$, the messages are thus either $\pm y_i$ or $\pm y_i(-1)^{x_i}$. In general, when $x_i$ takes $l_i$ values, there are $2^{l_i}$ possible messages for each partner to transmit. The choice of the optimal set of messages depends on a specific sign function. If it is weakly dependent on  $x_i$'s, say, for $N$ partners it is equal to $-1$ for only one $\vec{x}$.

The quantum protocol requires some additional assumptions. Let there be function $g(\vec{x})$ such that $f(v_{x})=Sign(g(\vec{x}))$ and $\frac{|g(\vec{x})|}{\sum_{\vec{x}}{|g(\vec{x})|}}$ be a probability that $\vec{x}$ is distributed among the partners. Additionally, let us assume that $\sum_{\vec{x}}g(\vec{x})E(\vec{x})=Q\leq B$, ($E(\vec{x})=\langle O_A(x_A)O_B(x_B)...\rangle$ denotes a quantum mechanical mean value of a tensor product of local observables $O_A(x_A),O_B(x_B),...$ with outcomes $\pm 1$) represent a Bell-type inequality violated by some state. Let the partners  happen to share some copies of that state. In quantum protocol, the partner basically conduct measurements $O_i(x_i)$ obtaining result $o^M_i$, $M$ enumerating the copy of a state used. The message is simply $m_i=y_io_i$, and the decider's answer is  $\prod_{i=A,B,...}m_i$. The overall probability that the answer is correct is
\begin{equation}
\label{pqu}
P_{QM}=\frac{1}{2}\left(1+\frac{Q}{\sum_{\vec{x}}|g(\vec{x})|}\right)
\end{equation}
In the classical protocol, the messages must be subject to Local Realism, hence the probability of success reads
\begin{equation}
P_{CL}\leq\frac{1}{2}\left(1+\frac{B}{\sum_{\vec{x}}|g(\vec{x})|}\right)
\end{equation}
(please keep in mind that the inequality is violated, i. e. $Q>B$).

Finally, let us explicitly demonstrate this advantage for a five-qubit  state with no five-partite correlations. The inequality we intend to use is
\begin{widetext}
\begin{eqnarray}
\frac{1}{16}&(A_1B_1C_1D_1-A_1B_1C_1D_2+3A_1B_1C_2D_1-3A_1B_1C_2D_2&\nonumber\\
&+A_1B_2C_1D_1-A_1B_2C_1D_2+3A_1B_2C_2D_1-3A_1B_2C_2D_2&\nonumber\\
&+A_2B_1C_1D_1-A_2B_1C_1D_2-A_2B_1C_2D_1+A_2B_1C_2D_2&\nonumber\\
&+A_2B_2C_1D_1-A_2B_2C_1D_2-A_2B_2C_2D_1+A_2B_2C_2D_2&\nonumber\\
+&B_1C_1D_1E_1-B_1C_1D_1E_2+3B_1C_1D_2E_1-3B_1C_1D_2E_2&\nonumber\\
&+B_1C_2D_1E_1-B_1C_2D_1E_2+3B_1C_2D_2E_1-3B_1C_2D_2E_2&\nonumber\\
&+B_2C_1D_1E_1-B_2C_1D_1E_2-B_2C_1D_2E_1+B_2C_1D_2E_2&\nonumber\\
&+B_2C_2D_1E_1-B_2C_2D_1E_2-B_2C_2D_2E_1+B_2C_2D_2E_2&\nonumber\\
+&C_1D_1E_1A_1-C_1D_1E_1A_2+3C_1D_1E_2A_1-3C_1D_1E_2A_2&\nonumber\\
&+C_1D_2E_1A_1-C_1D_2E_1A_2+3C_1D_2E_2A_1-3C_1D_2E_2A_2&\nonumber\\
&+C_2D_1E_1A_1-C_2D_1E_1A_2-C_2D_1E_2A_1+C_2D_1E_2A_2&\nonumber\\
&+C_2D_2E_1A_1-C_2D_2E_1A_2-C_2D_2E_2A_1+C_2D_2E_2A_2&\nonumber\\
+&D_1E_1A_1B_1-D_1E_1A_1B_2+3D_1E_1A_2B_1-3D_1E_1A_2B_2&\nonumber\\
&+D_1E_2A_1B_1-D_1E_2A_1B_2+3D_1E_2A_2B_1-3D_1E_2A_2B_2&\nonumber\\
&+D_2E_1A_1B_1-D_2E_1A_1B_2-D_2E_1A_2B_1+D_2E_1A_2B_2&\nonumber\\
&+D_2E_2A_1B_1-D_2E_2A_1B_2-D_2E_2A_2B_1+D_2E_2A_2B_2&\nonumber\\
&+E_1A_1B_1C_1-E_1A_1B_1C_2+3E_1A_1B_2C_1-3E_1A_1B_2C_2&\nonumber\\
&+E_1A_2B_1C_1-E_1A_2B_1C_2+3E_1A_2B_2C_1-3E_1A_2B_2C_2&\nonumber\\
&+E_2A_1B_1C_1-E_2A_1B_1C_2-E_2A_1B_2C_1+E_2A_1B_2C_2&\nonumber\\
&+E_2A_2B_1C_1-E_2A_2B_1C_2-E_2A_2B_2C_1+E_2A_2B_2C_2)&\leq 1.
\label{INEQ}
\end{eqnarray}
\end{widetext}
We have introduced a short-hand notation, $O_A(0)=1$ and  $A_{x_A}=\equiv O_A(x_A)(x_A=1,2)$, and similarly for other observers. Now the partner is marked with the label of an observable, rather than its position in the product. Notice that this inequality is invariant under cyclic permutations of the observers, as well as the fact that it utilizes only four-partite correlations. This means that it has doubly degenerate eigenvalues, and the equal mixture of the corresponding eigenstates does not exhibit any correlations between odd numbers of qubits. For simplicity, we will consider only one of these states keeping in mind that the other one has all relevant correlations equal.

The sign function to be computed can be extracted from Eqn. (\ref{INEQ}) to be
\begin{eqnarray}
f(\vec{x})=&q(x_E)(-1)^{x_A(1-x_D)+(1-x_A)(x_C(1-x_D)+(1-x_C)x_D)}&\nonumber\\
+&q(x_D)(-1)^{x_E(1-x_C)+(1-x_E)(x_B(1-x_C)+(1-x_B)x_E)}&\nonumber\\
+&q(x_C)(-1)^{x_D(1-x_B)+(1-x_D)(x_A(1-x_B)+(1-x_A)x_D)}&\nonumber\\
+&q(x_B)(-1)^{x_C(1-x_A)+(1-x_C)(x_E(1-x_A)+(1-x_E)x_C)}&\nonumber\\
+&q(x_A)(-1)^{x_B(1-x_E)+(1-x_B)(x_D(1-x_E)+(1-x_D)x_B)},
\end{eqnarray}
with $q(x)=1-\frac{1}{2}(x(3-x))$. An individual term of the function survives only if the corresponding $x_i$ is equal to 0, or in other words, when one of the observers is told not to conduct any measurement. Notice that from Eq. (\ref{INEQ}) it follows that only one of five partners receives $x_i=0$ hence indeed $f(\vec{x})=\pm 1$. This structure can be straight-forwardly extended to more values of $x_i$'s.
 
As we have mentioned the partners can send 8 possible classical messages: two containing no information about the value of $x_i$ whatsoever, $m_i=\pm y_i$, and three pairs distinguishing one of values of $x_i$, $m_i=\pm y_i(-1)^{\delta(x_i,j)}$. $\delta(\cdot,\cdot)$ stands for the Kronecker delta and $j=0,1,2$. Surprisingly, out of all possible classical strategies, the symmetric ones, in which partners share a common strategy of messaging, give the probability of success equal to $\frac{1}{2}$. However, there is a large class of strategies, for which the success probability is $\frac{17}{30}$. One of them is is that four partners send simply $y_i$ and the fifth (say, $E$) sends $y_E$ if $x_E=1$ and $-y_E$ otherwise. 

Let us now consider the quantum case. If the the observables at the disposal of each observer are $\sigma_x=\left(\begin{array}{cc}0&1\\1&0\end{array}\right)$ and $\sigma_z=\left(\begin{array}{cc}1&0\\0&-1\end{array}\right)$, there are two states associated with the maximal eigenvalue, $\frac{1}{4}+\sqrt{\frac{11}3}\cos\left(\frac{1}{3}\text{arccot}\sqrt{\frac{108}{1223}}\right)\approx 1.8086$. The equal mixture of these two eigenstates possesses no correlations between any odd number of particles. The partner, which gets $x_i=0$ simply sends $y_i$, the rest conduct measurement indicated by the value of their $x_i$'s, and send $y_io_i^M$. According to formula (\ref{pqu}) the probability of success is $0.620$.

Notice that we can purify the two states giving the maximal violation of the inequality. Then, the auxiliary party who holds the purifying ancilla shares entanglement with the five partners. Nevertheless, in the protocol they expoit only these correlations, which are unaccessible and unchangable by the external party.

%One may consider an alternative quantum scenario. Imagine that the partner, which gets $x_i=0$, disobeys this order and instead measures one on observables he would measure anyhow. Then, all the partners send $m_i=y_io_i^M$. The decider gives an answer, which is obtained in the majority voting between $m_Am_Bm_Cm_D$, $m_Am_Bm_Cm_E$, $m_Am_Bm_Dm_E$, $m_Am_Cm_Dm_E,$ and $m_Bm_Cm_Dm_E$.  A very brief inspection shows, however, that such a strategy cannot work. Note that every vote ignores one $y_i$, which is completely random. Even in the absurd situation, where all $y_i4$'s are determine, the majority voting strategy gives the success probability 0.5 for the discussed inequality.
 
Notice that the same protocol applies for even more complicated sing functions, even those which do not have the structure of a product of $y_i$'s. Instead we can consider 
\begin{eqnarray}
&f''(\vec{x},\vec{y})&\nonumber\\
=&\sum_{i=A,B,C,...}\left(\prod_{j=A,B,C,...}y_j\right)y_iq(x_i)f_i(\vec{x}),&\nonumber\\
&\frac{\partial f_i(\vec{x})}{\partial x_i}=0.
\end{eqnarray}
In such a case, the partner who gets $x_i=0$ simply sends 1.

In conclusion, we have demonstrated the instance of communication complexity reduction protocol, which utilizes only subcorrelations. This is a curious result, as the partners are able to improve their performace in jointly computing a certain function, even with one of them not sharing any useful data. It should be stressed that although we never utilize five-partite correlations, such a great advantage of the quantum protocol over the classical one is due to genuine  multipartite entanglement.

It is also worth stressing that we are no longer limited to functions, which are proportional to products of all $y_i$'s, but we can also consider sign functions, which are sums of few terms, each dependent on a different subset of the bits. Each such term would be activated by a specific combination of $x_i$'s. This relaxation contributes to the significance and of the problem.

The example shows a distinction between inequalities with subcorrelations and those with many settings per each side. Although some inequalities from both classes are strongly linked with one another (by substituting local some measurements with the unit operator.), they differ in problems they allow to solve. In case of using subcorrelations we are able to increase the fidelity of computing functions with non-product structure.

One important aspect is that we believe that $\vec{x}$'s are distributed according to a certain probability distribution. We also assume that detection/non-detection of photons in each lab is a stochastic, independent process. Hence it is impossible to guarantee that all qubits, on which the partners need to conduct measurements, must be detected. Even though we are not observing full correlations, we need to be confident about detecting a particle in each laboratory in every run of the experiment. 

This work is a part of IDEAS PLUS Grant (IdP211 000361) from the Ministry of Sciencie of The Higher Education of Poland. The Author  acknowledges the support from the Foundation for Polish Science (Program HOMING PLUS). This work is dedicated to Prof. Marek \.Zukowski on the occasion of his 60th birthday.% and thanks Daniel Richart for useful remarks.

\end{document}